# TRANSPIRATION- AND PRECIPITATION-INDUCED SUBSURFACE WATER FLOW OBSERVED USING THE SELF-POTENTIAL METHOD

Emily B. Voytek*[1], Holly R. Barnard[2], Damien Jougnot[3], and Kamini Singha[1]

*Corresponding author, now at Institute of Earth Sciences, University of Lausanne, 1015 Lausanne, Switzerland, emily.voytek@unil.ch

[1]Hydrologic Science and Engineering Program, Colorado School of Mines, Golden, Colorado.

[2]Institute of Arctic and Alpine Research, Department of Geography, University of Colorado Boulder, Boulder, CO 80309, USA

[3]Sorbonne Université, CNRS, EPHE, UMR 7619 METIS, Paris, France

**CORE IDEAS**

- Self-potential (SP) measurements are sensitive to transpiration- and precipitation-induced water movement
- Measured SP data and pressure gradients suggest contrasting directions of flow
- Daytime increases in sapflow correspond with reduced rates of downward flow in SP measurements
- Coupled fluid flow and electrical models supports transpiration origin of observed diel fluctuations in SP signals
- Diel fluctuations in soil-water movement disappear with tree dormancy; precipitation-induced infiltration dominates

**ABSTRACT**

Movement of soil moisture associated with tree root-water uptake is ecologically important, but technically challenging to measure. Here, the self-potential (SP) method, a passive electrical geophysical method, is used to characterize water flow in situ. Unlike tensiometers, which use a measurement of state (i.e. matric pressure) at two locations to infer fluid flow, the SP method directly measures signals generated by water movement. We collected SP measurements in a two-dimensional array at the base of a Douglas-fir tree (*Pseudotsuga menziesii*) in the H.J. Andrews Experimental Forest in western Oregon over five months to provide insight on the propagation of transpiration signals into the subsurface under variable soil moisture. During dry conditions, SP data appear to show downward unsaturated flow, while nearby tensiometer data appear to suggest upward flow during this period. After the trees enter dormancy in the fall, precipitation-induced vertical flow dominates in the SP and tensiometer data. Diel variations in SP data correspond to periods of tree transpiration. Changes in volumetric water content occurring from soil moisture movement during transpiration are not large enough to appear in volumetric water content data. Fluid flow and electrokinetic coupling (i.e. electrical potential distribution) were simulated using COMSOL Multiphysics to explore the system controls on field data. The coupled model, which included a root-water uptake term, reproduced components of both the long-term and diel variations in SP measurements, thus indicating that SP has potential to provide spatially and temporally dense measurements of transpiration-induced changes in water flow. This manuscript presents the first SP measurements focusing on the movement of soil moisture in response to tree transpiration.

# 1. Introduction

Quantifying the interconnection between evapotranspiration (ET) and groundwater is a necessary, but technically challenging, component of evaluating critical zone processes. The links between hydrology and plant ecology are particularly complex in the vadose zone, where soil moisture availability controls vegetation distribution, and the vegetation affects the soil moisture distribution (e.g., Tromp-van Meerveld and McDonnell, 2006, D'Odorico et al., 2007; Moore et al., 2011; Swetnam et al., 2017;). Vadose zone processes within hillslopes, such as plant-water uptake, mediate groundwater discharge to streams (e.g., Asbjornsen et al., 2011). In temperate climates, transpiration has been implicated in producing diel fluctuations in streamflow, particularly during baseflow conditions when transpiration rates are high relative to total streamflow in smaller streams (e.g., Bond et al., 2002; Burt, 1979; Graham et al., 2013; Lundquist and Cayan, 2002). The dominant explanations for the transpiration-driven origin of these diel fluctuations are: (1) vegetation uptake of water from lateral subsurface flowpaths linking hillslopes to streams (Bren, 1997), (2) removal of hyporheic water from stream-side aquifers by vegetation (Bond et al., 2002), or (3) a combination of both. The relative contribution and timing of the processes leading to diel streamflow variability, particularly under differing hydrologic regimes, remains poorly described with existing data (Bond et al., 2002; Graham et al., 2013; Newman et al., 2006; Voltz et al., 2013; Wondzell et al., 2010). In particular, the mechanisms which propagate signals from upslope processes, including transpiration, through hillslopes to streams remain difficult to parse (Ali et al., 2011; McCutcheon et al., 2017; McGuire and McDonnell, 2010). To better describe subsurface processes that link transpiration and streamflow and to identify related flowpaths within catchments, development of new sensors capable of measuring in situ water movement beyond the point scale (i.e. integrative measurements) is needed.

Here, we explore the use of the self-potential (SP) method, a passive electrical geophysical tool that is sensitive to water flow (in saturated and unsaturated conditions), to quantify subsurface processes occurring in response to plant transpiration. The SP method measures electrical potential differences generated by natural current-inducing processes in the ground, including water movement. SP has been used to measure unsaturated flow rates in one dimension (Doussan et al., 2002; Thony et al., 1997) and to measure sap flow within a tree trunk (Gibert et al., 2006), but not previously used to evaluate vegetation-induced water movement in the vadose zone. The objective of this work is therefore to: 1) evaluate the sensitivity of SP to detect the propagation of diel transpiration signals into the unsaturated subsurface and 2) explore the effect of soil moisture on propagation of transpiration signals at the single-tree scale at the H.J. Andrews Experimental Forest (HJA), Oregon, USA.

## 2. Background

### *2.1. Vegetation-hydrology interactions*

Many studies, reviewed in Hewlett and Hibbert (1967), Bosch and Hewlett (1982), and Gribovszki et al. (2010), suggest that diel patterns of transpiration are the prominent driver of streamflow fluctuations in forested systems; such patterns have been seen in streamflows since the 1930s (Blaney et al., 1933, 1930; Troxell, 1936; White, 1932). These connections have been tested through comparison studies of catchments with and without vegetation (e.g. Dunford and Fletcher, 1947; Rothacher, 1965) and time-series analysis of transpiration and resultant streamflow (e.g. Bond et al., 2002). Despite the strong coupling observed between transpiration and streamflow time series, isotopic work has found, in some systems, that stream water and water used by near-stream plants are not from the same source (e.g., Brooks et al., 2010). Isotopic data suggest that during dry seasons, when diel fluctuations in streamflow are highest, trees and other vegetation are using water that is disconnected from the groundwater discharging to the streams (Brooks et al., 2010; Dawson and Ehleringer, 1991; Evaristo et al., 2015). However, this hypothesis remains open for exploration in light of other data suggesting different mechanisms that may be responsible for the isotopic differences (e.g., McCutcheon et al., 2017; Vargas et al., 2017). The complexity of the subsurface and the limited data available in this part of the critical zone hamper evaluation of the mechanisms that propagate transpiration signals from trees, through the hillslope, to streams.

Methods used to quantitatively determine which hillslope processes are occurring in the subsurface generally include catchment-averaged or point-scale measurements, with few support volumes in between. Catchment-averaged measurements include analysis of streamflow volume or stream-water chemistry, for example. Point-scale measurements include data like soil moisture content and matric potential. However, due to sensor sensitivity, changes in soil moisture associated with plant-water use may be limited to periods when soil moisture is high and/or plant uptake is large (e.g., Musters et al., 2000). Geophysical methods may provide spatially integrative measurements between these scales that can be useful in evaluating hillslope scale processes related to plant-water use and the impact on subsurface water flow. For example, electrical resistivity has been used to map soil moisture content changes associated with vegetation (e.g. Jayawickreme et al., 2010, 2008; Robinson et al., 2012; al Hagrey, 2006; Mares et al., 2016; Bass et al., 2017). Electromagnetic induction has also been used to map soil type and water content (e.g. Doolittle and Brevik, 2014). However, unlike other hydrologic or geophysical methods, which are sensitive to static or state variables (e.g. water content, lithology), SP is sensitive to dynamic processes (e.g. water flow, ionic fluxes, electron transfer). For this reason, use of SP in hydrologic studies has increased in recent years (e.g. Darnet and Marquis, 2004; Jougnot et al., 2015; Linde et al., 2011; Maineult et al., 2008).

*2.2. SP to Map Critical Zone Processes*

The growing use of SP in hydrologic studies includes the use of SP to analyze critical zone processes. Thony et al. (1997) and Doussan et al. (2002) used SP to monitor vertical movement of newly infiltrated water from precipitation events. These studies showed that SP was capable of estimating water flux in the vertical direction on the scales of decimeters (resolution is electrode-spacing dependent), but interpretation of signal magnitudes in response to precipitation events required the use of a site-specific coupling coefficient. Darnet and Marquis (2004) refined the interpretation of SP data in vadose-zone processes through the inclusion of a previously omitted saturation-dependent coupling coefficient to explain the magnitude of signal generation, which accounted for some site-specific differences observed in the earlier work.

Note that in ecohydrology, where the overall quantity of water in the soil is important, soils are often described by their soil moisture, or volumetric water content (VWC). However, in discussion of the physics of SP, including in the following background sections, the fraction of pore space that is filled is more pertinent, so saturation, $S_w$ [-], is used. Saturation is related to VWC [-] by porosity, $n$ [-]:

$$VWC = S_w \, n \,. \tag{1}$$

The theorized saturation dependence of SP signal strength has since been confirmed and expanded on in many additional theoretical and laboratory studies (Guichet et al., 2003; Revil and Cerepi, 2004; Vinogradov and Jackson, 2011), but each of these examples assumes near-uniform one-dimensional flow. As described in more detail in the background and discussion below, the amplitude of measured SP signals (voltages), is a function of the strength of current source (e.g. water movement) and the ability of current to propagate within the material (i.e.

electrical conductivity, which is dependent on water content and ionic strength of pore water). In this work, we build on the existing use of SP to evaluate near-vertical (one-dimensional) soil moisture movement in response to infiltration and evapotranspiration (e.g. Sailhac et al., 2004) to include multidimensional soil-water movement associated with tree root-water uptake, and also discuss some of the challenges of using this method in this manner.

*2.3. SP Background*

SP relies on measurements of electrical potential differences [V] generated by natural currents in the ground. The amplitude of measured SP voltages depends on the magnitude of currents generated and the electrical conductivity of the ground material as described by a generalized version of Ohm's law in the framework proposed by Sill (1983):

$$\boldsymbol{J} = \sigma \boldsymbol{E} + \boldsymbol{J}_S \tag{2}$$

where $\boldsymbol{J}$ is the macroscopic current density [A m$^{-2}$], $\sigma$ is the electrical conductivity of the ground [S m$^{-1}$], $\boldsymbol{E}$ is the electrical field [V m$^{-1}$] and $\boldsymbol{J}_S$ is the source current density [A m$^{-2}$].

The first term, $\sigma \boldsymbol{E}$, describes the conduction current density, or how electrical signals propagate through the material, while $\boldsymbol{J}_S$ describes the distribution of currents generated by water movement or other possible current sources, discussed below. The electrical field is further defined by:

$$\boldsymbol{E} = -\nabla V \tag{3}$$

where $V$ is the electrical potential [V], measured in the SP method. The SP signal is the electrical potential difference between a reference and a potential electrode:

$$SP_i = V_i - V_{ref}. \tag{4}$$

To solve Equation 2 for the SP voltages, the above constitutive equations describing current density must be combined with a charge conservation equation, which describes the current quantity. At the quasi-static limit of Maxwell's equations, the conservation of charge is described by:

$$\nabla \cdot \mathbf{J} = 0. \tag{5}$$

While charges can move, there is not source or sink of changes. Combining Equations 2, 4 and 5 produces the field equation describing the complete electrical problem:

$$\nabla \cdot (\sigma \nabla V) = \nabla \cdot \mathbf{J}_S. \tag{6}$$

Multiple processes, including ground- or soil-water movement, thermal and chemical diffusion (e.g. Leinov and Jackson, 2014), and under specific conditions, redox gradients (Hubbard et al., 2011), can generate electrical currents and contribute to $\mathbf{J}_S$ (Revil and Jardani, 2013).

Prior to interpretation of SP data, corrections must be made to the data to account for equipment drift and other sources of error. For example, temperature variations between the reference and measuring electrodes can result in undesired voltage changes in SP measurements. Petiau-type electrodes (Petiau, 2000) have improved temperature stability as compared to electrodes used previously, but measurements must still be corrected for temperature differences between the two electrodes:

$$SP_i^T = \alpha(T_i - T_{ref}) \tag{7}$$

where $SP_i^T$ is the temperature correction to be applied, $\alpha$ is the electrode-type correction factor (0.2 mV/°C for Petiau-type; Petiau, 2000), and $T_i$ and $T_{ref}$ are temperatures at the measurement and reference electrodes respectively. SP measurements are also sensitive to temporally variable electrode drift caused by electrode age and changing chemistry in the immediate vicinity of the electrode (Jougnot and Linde, 2013). In surface SP surveys, this electrode drift is corrected by measuring the voltage difference between the two electrodes through time by placing them in direct contact with each other on occasion during data collection. In long-term measurements where the electrodes are buried in the ground, this type of correction is not possible, but we estimate the effect to be less than 0.2 mV/month based on manufacturer's data (SDEC, Reignac sur Indre, France).

The primary signal of interest in this work is the current generated by the movement of water in the vadose zone (i.e. the electrokinetic phenomenon). Water movement produces electrical current due the existence of the electrical double layer (EDL) at the pore water-mineral interface. The EDL develops when a mineral surface is in contact with water, which alters the surface charge of the mineral and the surrounding water chemistry. For example, under near-neutral pH conditions, 5-8, the surface charge of a silica grain is negative (Revil and Jardani, 2013). The charged surface attracts ions of the opposing charge (positive in the case of silica) from the bulk pore water, the so-called counter-ions. Certain counter-ions sorb directly onto the mineral surface, forming the less-mobile Stern layer on the surface of the grain, while additional counter-ions exist in the more-mobile diffuse layer (Figure 1). The diffuse layer exists as ions are simultaneously attracted to the excess surface charge of the mineral and repelled by the enriched concentration of like charges (Revil and Jardani, 2013). Under saturated conditions, the net charge of the EDL, which comprises both the less-mobile Stern layer and more-mobile diffuse layer, is positive relative to the surrounding pore water and is called the total excess charge density, $Q_v$ [C m$^{-3}$]. $Q_v$ is dependent on soil properties (mineralogy, pH, and pore-water chemistry) and is related to the cation-exchange capacity (CEC, [meq L$^{-1}$]) through:

$$Q_v = \rho_g \left(\frac{1-n}{n}\right) CEC \qquad (8)$$

where $\rho_g$ is the grain density [kg m$^{-3}$] and $n$ is porosity [-] (Waxman and Smits, 1968). CEC measures all of the excess charges in the EDL, mobile and immobile; however, current is only generated by moving charges. Consequently, only the excess charge in the mobile, diffuse layer in the EDL contributes to generation of SP signals. This value, the *effective* excess charge density, $\hat{Q}_v$ [C m$^{-3}$], can be 3-4 orders of magnitude smaller than total excess charge (Jougnot et al., 2012).

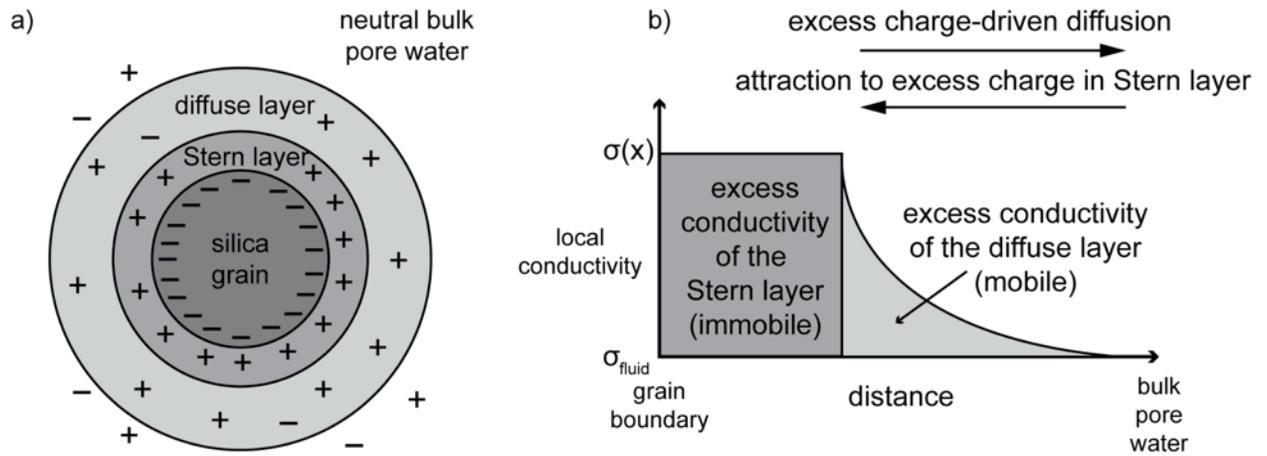

*Figure 1. a) Schematic of electrical double layer (EDL) formed on the exterior of a mineral grain when in contact with water. b) Plot of relative charges in the pore water with relative distance from the mineral surface. Modified from Revil and Jardani (2013).*

Determining $\hat{Q}_v$ from $Q_v$ is difficult; therefore, constitutive relationships between $\hat{Q}_v$ and other measureable parameters have been developed. Jardani et al. (2007) presents an empirical relationship between $\hat{Q}_v$ and permeability, $k$ [m$^2$] defined for saturated conditions:

$$log(\hat{Q}_v) = -9.2349 - 0.8219 log(k). \tag{9}$$

$\hat{Q}_v$ is dependent on the level of saturation, or percentage of filled pore space (see discussion of Equation 1). To account for the effect of saturation on electrokinetic coupling, Linde et al. (2007) and Revil et al. (2007) proposed a theoretical framework using a volume-averaging upscaling procedure. This upscaling procedure describes the effect of water saturation on the corresponding effective excess charge density:

$$\hat{Q}_v(S_w) = \frac{\hat{Q}_{v,sat}}{S_w} \qquad (10)$$

where $\hat{Q}_v(S_w)$ [C m$^{-3}$] is the effective excess charge as a function of saturation. As the saturation in the medium decreases, the quantity of excess surface charges remains constant, and therefore $\hat{Q}_v$, a charge density, increases. Although the volume-averaging approach of Linde et al. (2007) provides a first-order approximation of $\hat{Q}_v$ in homogeneous media at high saturation, the approach does not reproduce SP signal amplitudes in low-saturation soils (Jougnot et al., 2012; 2015; Allègre et al. 2014). To better fit signals produced by water movement in soils at low saturations, Jougnot et al. (2012) proposed a "flux-averaging" upscaling approach (i.e., an averaging procedure based on the water flux distribution in the soil). In this approach, the soil is treated as a bundle of capillaries of varying size. Depending on the saturation, which in this model is related to matric potential by a specified soil water retention function, the individual capillaries are defined as either saturated or dry. Dry capillaries do not contribute to the effective excess charge of a soil. The contribution of individual saturated capillaries is related to the radius-dependent distribution of pore water velocity and excess charges in the diffuse layer. The soil-specific effective excess charge is then calculated by integrating the effective excess charge of saturated capillaries for given pore-size distribution:

$$\hat{Q}_v(S_w) = \frac{\int_{R_{min}}^{R_{S_w}} \hat{Q}_v^R(R) v(R) f_D(R) dR}{\int_{R_{min}}^{R_{S_w}} v(R) f_D(R) dR} \qquad (11)$$

where $R$ [m] is the capillary size, $f_D$ [-] is the equivalent pore-size distribution function inferred from hydrodynamic parameters (i.e. van Genuchten parameters and permeability $k$) and $v_R$ [m s$^{-1}$] is the mean pore-water velocity for a given capillary radius. This approach is more complex than Equation 10 but better describes SP amplitudes in natural media over a large range of saturations in which $\hat{Q}_v$ can increase multiple orders of magnitude as saturation decreases (Jougnot et al., 2012; 2015, Zhang et al., 2017; Soldi et al., 2019). Given the large variation in saturation observed during the period of data collection in this work, we use the Jougnot et al. (2012) relative permeability (RP) flux-averaging approach in this work (Figure 2).

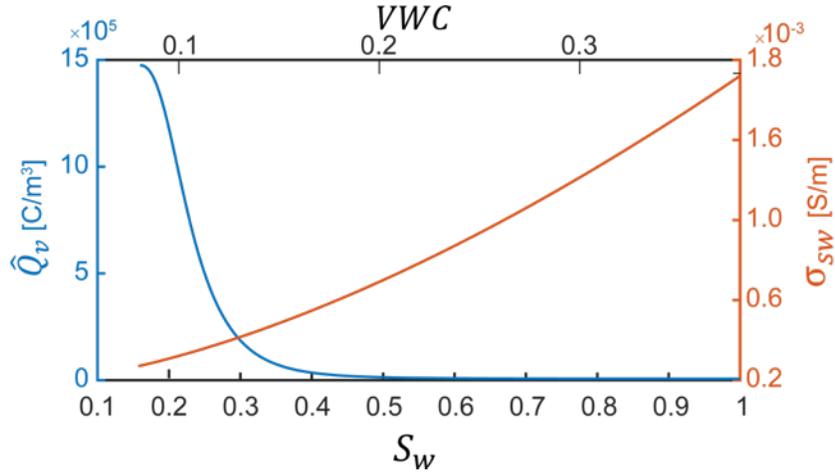

*Figure 2. Saturation-dependent effective excess charge, $\hat{Q}_{v,sw}$, and effective conductivity accounting for saturation, $\sigma_{sw}$. $S_w$: saturation, VWC: volumetric water content.*

Once $\hat{Q}_v(S_w)$ is defined, the streaming-current term of water movement in variable saturated conditions can be written:

$$\boldsymbol{J}_{streaming} = \hat{Q}_v(S_w)\boldsymbol{U} \qquad (12)$$

where **U** is the soil moisture velocity [m s$^{-1}$]. Soil moisture velocity can be solved according to a modified version of Richards equation:

$$-\nabla \cdot \boldsymbol{U} = \nabla \cdot [K(S_w)\nabla h] \qquad (13)$$

where $K(S_w)$ is saturation-dependent hydraulic conductivity [m s$^{-1}$], and $h$ is the total matric head [m].

In addition to the previously discussed saturation dependence of $\hat{Q}_v$, ground electrical conductivity, σ, is a saturation-dependent parameter that increases with added soil moisture and is needed to estimate electrical flow (Equation 6). Under certain conditions this relationship should obey Archie's law (Archie, 1942), but this equation does not account for surface conductivity present in many natural soils. Instead, a modification of Archie's law obtained through volume averaging and that includes a surface conductivity term not scaled by saturation as in the Waxman and Smits model (1968) (Linde et al., 2006; based on Pride, 1994) has been found to best predict electrical conductivity of loamy soils (Laloy et al., 2011):

$$\sigma_{sw} = \frac{1}{F}\left[S_w^b \sigma_f + (F-1)\sigma_s\right] \qquad (14)$$

where $\sigma_{sw}$ is the effective conductivity with varying saturation [S m$^{-1}$]; $b$ is Archie's second, or saturation, exponent [-] (e.g. Waxman and Smits, 1968); $F$ is the formation factor [-]; $\sigma_f$ is the fluid conductivity [S m$^{-1}$]; and $\sigma_s$ is the surface conductivity [S m$^{-1}$].

In variably saturated conditions, Equations 6 and 12 can be written as:

$$\nabla \cdot (\sigma_{sw} \nabla V) = \nabla \cdot (\hat{Q}_v(S_w)\boldsymbol{U}) . \qquad (15)$$

Using this equation, given information on electrical conductivity ($\sigma_{sw}$) and effective excess charge ($\hat{Q}_v$), we can use measured SP values to evaluate relative unsaturated flow rates through time and therefore examine the role of trees in subsurface water redistribution. In one-dimensional flow through homogenous material, under near-neutral pH conditions (5-8), voltages generally increase in the direction of flow. However, when considering three-dimensional flow and heterogeneous electrical conductivity due to changing moisture content, such as exists in the work presented here, systems are too complex for this simple relation to hold true. It is consequently necessary to numerically model the system to confirm the directional interpretation of self-potential data.

## 3. Field Site and Methods

Data were collected in Watershed 10 (WS10), a 0.1-km$^2$ watershed of the H.J. Andrews (HJA) Experimental Forest in the western Cascade Mountains between June and November 2016. The steep catchment (27-48° slopes) ranges in elevation from 480 m to 565 m (McGuire et al., 2007) and is underlain by highly weathered andesitic tuffs and coarse breccias. The soils are residual and colluvial deposits (mesic Andic Humudepts; Soil Survey Staff, accessed 04/15/2017) with an average depth of 1.3 m. Soil textures in the upper meter are gravelly, silty-clay loams to very gravelly clay loams, with slightly blockier textures below 0.7 m (Harr, 1977). The soils cover weathered saprolite that can be up to 7 m thick (3.6 m average) (Harr and McCorison, 1979) and the depth to groundwater is approximately 3.5 m near our sampling site (Gabriellei et al., 2012). The mean precipitation at the HJA is 2220 mm, with approximately 80% falling between October and April (McGuire et al., 2007). The catchment was clear cut in 1975 to evaluate the effects of local forestry practices on the catchment (Harr, 1977). Vegetation is now dominated by ~40 year old Douglas-fir (*Pseudotsuga menziesii*) forest.

### *3.1. Tree Selection*

We selected a single Douglas-fir tree on which to focus our SP and corroboratory measurements; selection of the tree was based on ease of access and proximity to other ongoing experiments. We selected a relatively isolated Douglas-fir tree (23.3 cm diameter at breast height, dbh), but it was not possible to exclude the effects of other vegetation from our measurements as root zones of Douglas-fir trees are known to extend 3 - 4 times maximum canopy width (e.g. Eis, 1987). Additionally, dense understory vegetation was present, including western swordfern (*Polystichum munitum*), deer fern (*Blechnum spicant*), and Oregon grape (*Berberis nervosa*). The nearest neighboring trees were two other Douglas-fir trees located 3.2 m downslope (30.7 cm dbh) and 4.3 m upslope (23 cm dbh). The stream draining the catchment was approximately 11 m downslope from the selected tree.

### *3.2. Self-potential measurements*

We collected SP data in a two-dimensional subsurface array at the base of the Douglas-fir tree (Figure 3a). The array consisted of four Petiau-type non-polarizing electrodes (Petiau, 2000) on the downslope side of the tree. The array included two electrodes at 0.3 m and two electrodes at 0.8 m depth (measured to the porous tip of the electrode). This configuration provided for measurements of SP between four pairs of electrodes, which will be referred to by their positions relative to the tree and ground surface: two vertical pairs (inner, 0.1 m from the tree; and outer, 0.9 m from the tree), and two horizontal pairs (upper, 0.3 m below the ground surface; and lower, 0.8 m below the ground surface; Figure 3a) For the vertical electrode pairs, inner and outer, the lower electrode is taken to be the reference such that positive values suggest upward movement. For the horizontal pairs, upper and lower, the reference is the outer electrode such that positive values suggest inward movement (towards tree axis; Figure 3a).

Electrodes were installed by hand augering to the desired depths, inserting a small amount of bentonite (~0.5 L), emplacing the electrodes, and refilling the hole with the native soil. The bentonite was added to ensure good contact between the electrodes and the ground and to reduce the chance of electrodes drying out. Prior to installation, the electrodes were checked using a handheld Fluke 87V voltmeter (accuracy: ±0.1 mV) to ensure the tip-to-tip differential voltage between all pairs was <1 mV. Tip-to-tip differences greater than 1 mV can indicate electrode failure such as poor internal wire connections or drying out of the internal solution. The electrodes were left in place for the duration of data collection (4 months), reducing position uncertainty. Voltages were recorded between four pairs of electrodes using a Campbell Scientific CR1000 logger at 15-min intervals (resolution: 0.67 mV, accuracy: ±2 mV for 2.5 V range, internal impedance: 20 GΩ). After collection, the data were corrected for the known temperature drift of 0.2 mV/°C in the Petiau-type electrodes (Petiau, 2000; Equation 7) based on ground temperature measurements, described below.

The addition of a solar panel to the data logger on July 11 resulted in undesired electrical noise during daylight hours. These noisy data have been filtered from the plot of the full record using a standard deviation filter (Figure 4e), as they obscured detection the discrete event signals discussed in the text. Close-ups of these data for July 5-8 (before solar panel; Figure 5e) and September 8-11 (after solar panel; Figure 5j) are presented for comparison.

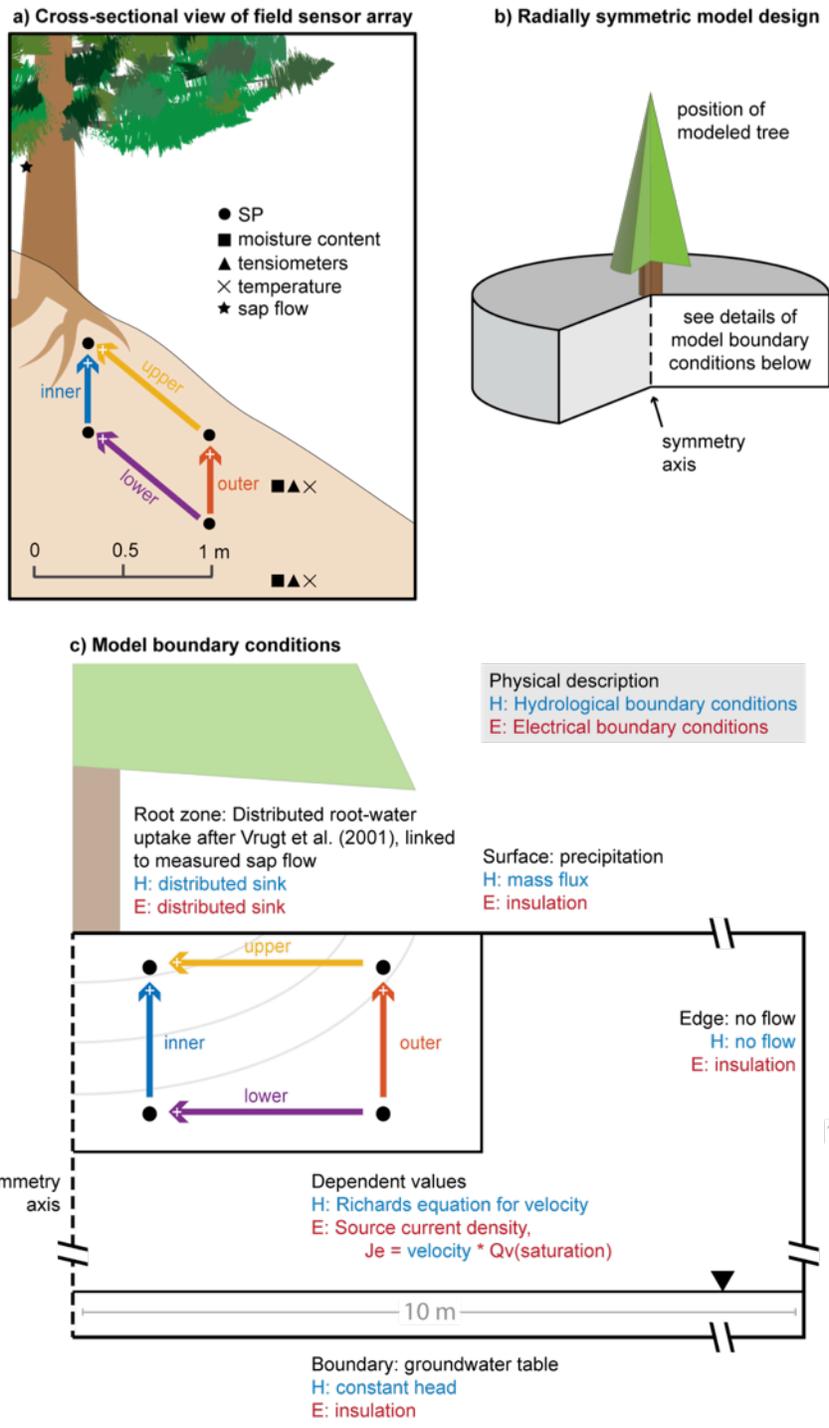

*Figure 3 a) Cross-sectional view of sensor array relative to selected tree. Four SP electrodes were installed in a grid at 0.3 and 0.8 m depth, and 0.1 m and 0.9 m away from the tree. Two tensiometers, two soil moisture sensors and two thermocouples were installed 0.3 m downslope from the lower SP electrodes. b) Example of radially symmetric model c) Conceptual model used to investigate SP signals associated with root-water uptake and description of hydrological and electrical boundary conditions in the numerical model. Arrows indicate dipole orientation: the reference electrode is located at the flat end and measurement electrode at arrowhead.*

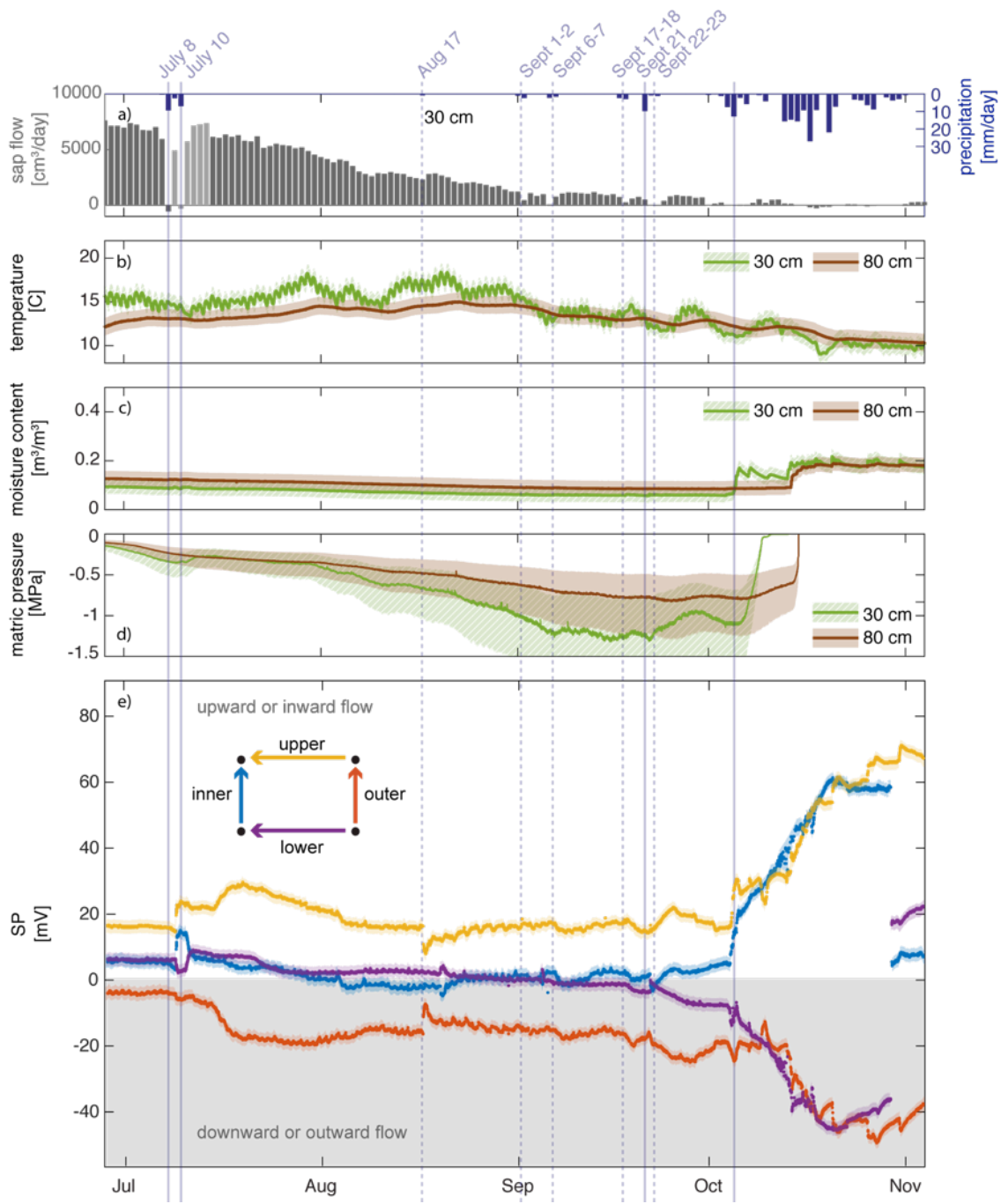

*Figure 4. a) Tree transpiration calculated from measured sap flow within the tree and precipitation recorded at the HJA PRIMET station. Missing July 9-14th sap flow data were infilled from measurements on nearby Douglas-fir trees due to equipment failure, b) ground temperature, c) soil moisture, d) matric potential and e) measured SP voltage differences between subsurface electrodes. Shaded areas are the manufacturer-reported accuracy of measurements, as discussed in the text. Vertical lines indicate the rainfall events discussed in the text. Solid lines are periods of >10 mm/day, dashed are 2-10 mm/day.*

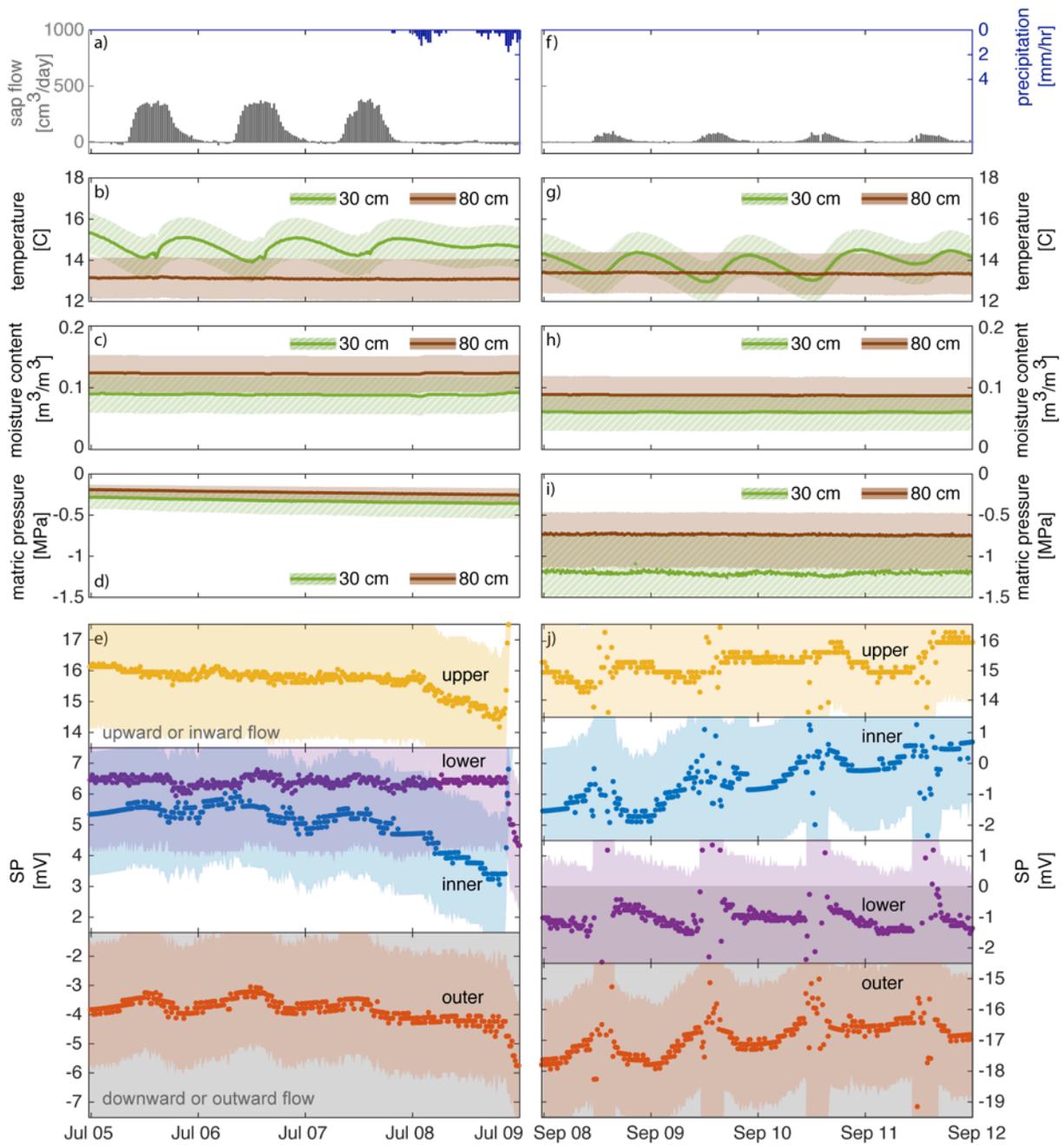

*Figure 5. Measured field data over four days from the full record: a-e) July 5-8 and g-j) September 8-11. a) and f) Tree transpiration calculated from measured sap flow and precipitation recorded at the HJA PriMet station, b) and g) ground temperature, c) and h) soil moisture, d) and i) matric potential and e) and j) measured SP voltage differences between each pair of subsurface electrodes. Shaded areas are the manufacturer-reported accuracy of measurements, as discussed in the text. Mid-day noise in the September SP data are due to addition of a solar panel to the field setup.*

### 3.3. Corroboratory measurements

In addition to SP measurements, soil moisture content, temperature, and matric potential were recorded at 15-min intervals in the immediate vicinity of the SP array (Figure 3a), and recorded on the Campbell Scientific CR1000. Soil moisture content was measured at 0.3 m and 0.8 m depths using two Decagon EC-5 sensors (Meter Group, resolution: 0.001 $m^3/m^3$, accuracy: ±0.03 $m^3/m^3$). Soil properties at the depths investigated were considered relatively homogenous based on previous soil surveys at the site (Harr, 1977) and no soil-specific calibrations were used for the EC-5 sensors. Soil temperature was recorded using three Type T thermocouples (accuracy: ±1 °C). Soil matric potentials were also measured at 0.3 m and 0.8 m below land surface, using Tensiomark sensors (ecoTech, Bonn, Germany; resolution: ±0.01 pF, accuracy: 5% of pF). The sensors report values in pF, which is related to MPa by pF = log10(pressure in hPa). At 1 MPa, a resolution of 0.01 pF corresponds to ±0.02 MPa, and the bounds of 5% pF accuracy are 0.45 and 2.24 MPa).

The matric potential sensors are independent of soil texture because they measure heat dissipation in the incorporated ceramic tip; consequently, no site-specific soil calibration was used. Above ground, heat-pulse sap flow sensors (Burgess et al., 2001) were installed 1.8 and 3.5 cm into the tree at 1.2 m above land surface and sap flow velocities were measured at 30-min intervals on a separate logger. Sap flow velocity was converted to tree transpiration rate using the methods of Dawson et al. (2007) and Hu et al. (2010). The relative positions of all sensors were measured using a Leica total station. In addition to data collected specifically for this project, precipitation data were used from the PRIMET station run by HJA, located at station headquarters (< 1 km for the field site).

## 4. Coupled Soil Water Flow and Electrical Modeling

A coupled fluid flow and electrical model was used to explore controls on the measured SP data. The model is 2-D with radial symmetry (Figure 3b), with the tree as the center axis. The modeled domain has a 10-m radius and a 10-m depth. The ground is assumed homogenous, and extends beyond the zone of influence of the tree (Figure 3a). The model was implemented and solved in COMSOL 5.3. The fluid flow and electrical models were run sequentially (i.e. electrokinetic semi-coupling as in Linde et al. (2011)), and the details of each are discussed in that order below. The model was calculated at 30-min time steps to coincide with the interval of the sap flow measurements.

In the model, unsaturated water flow was solved according to a modified Richards equation (Equation 13) using the van Genuchten parameterization (van Genuchten, 1980) of moisture-dependent hydraulic conductivity, $K$. The hydrologic boundary conditions are based on observed field conditions: the basal boundary is a constant head, based on observed groundwater levels (3.5 m below ground surface); the ground surface is a prescribed flux, linked to the measured precipitation at the field site (range: 0 – 54 mm/day; Figure 4a and Figure 5a) and the outside edge is a no-flow boundary, where the tree is assumed to have no influence. The porosity, hydraulic conductivity, and soil moisture retention parameters are based on measured values from Harr (1977) and Carsel and Parrish (1988) and are detailed in Table 1. The VWC to resistivity relationship used (Equation 14) is derived from these same parameters, and comparison with field electrical resistivity data from ongoing work at the site (R. Harmon, personal communication).

Root-water uptake was included as a distributed source/sink centered on the axis of symmetry following the model of Vrugt et al. (2001), which comprises two terms. The first term,

$\beta$, describes the geometric distribution of root-water uptake under unstressed conditions (Raats, 1974):

$$\beta(r,z) = \left[\left(1 - \frac{z}{z_m}\right)\right]\left[\left(1 - \frac{r}{r_m}\right)\right] e^{-\left(\frac{p_z}{z_m}|z^*-z| + \frac{p_r}{r_m}|r^*-r|\right)} \tag{16}$$

where $r$ [m] and $z$ [m] are the position relative to the base of the tree, $r_m$ [m] and $z_m$ [m] are the maximum radial rooting length and rooting depth, and $p_z$ [-], $z^*$ [m], $p_r$ [-], and $r^*$[m] are empirical parameters describing the distribution of uptake.

To account for changing moisture conditions, $\beta$ is combined with a soil-stress equation, defined by γ(r, z, h) [-], from van Genuchten (1987):

$$\gamma(r,z,h,t) = \frac{1}{\left[1 + \left(\frac{h(r,z,t)}{h_{50}}\right)^3\right]} \tag{17}$$

where $h$ is the total matric head [m] at a location ($r$,$z$), $h_{50}$ [m] is the soil-water pressure head at which root-water uptake is reduced by 50% (defined by mean cavitation pressure [P] in tree physiology literature; Table 1). For our Douglas-fir root model, an experimentally derived $h_{50}$ value of -200 m (-2 MPa; Sperry and Ikeda, 1997) was used. A maximum rooting depth of 1 m was used, corresponding to the observed thickness of soil surrounding the selected Douglas-fir tree. Curt et al. (2001) found little Douglas-fir root biomass in the substratum below soil, including in areas of fractured substratum, similar to the system at HJA.

The localized source/sink applied to our axially symmetric model can be calculated normalizing the product of the two terms, $\beta$ (Equation 16) and $\gamma$ (Equation 17):

$$Q(r,z,h,t) = \frac{\pi r_m^2 \, \beta(r,z)\, \gamma(r,z,h,t)}{2\pi \int_0^{z_m} \int_0^{r_m} \beta(r,z)\, \gamma(r,z,h,t)\, dr\, dz} T_{sap} \tag{18}$$

where $T_{sap}$ is total transpiration [m³s⁻¹] calculated from sap flow measurements (see Section 3.3).

Once the fluid flow problem is solved at each time step, the electrical problem is solved according to Ohm's law (Equation 15). The model has insulation conditions at all external boundaries, combined with an external current density linked to calculated flow velocities from the fluid flow model (Equation 12). In the electrical simulation, both the effective excess charge density, $\hat{Q}_v$ and bulk conductivity $\sigma_{sw}$ are saturation dependent as described in Equations 11 and 14 (Table 1 and Figure 2). From the resulting voltage distribution, voltage differences comparable to the electrode placement in the field are extracted for analysis.

*Table 1. Parameters used in coupled fluid flow and electrical models.*

|  | Variable | Value | Unit | Description |
|---|---|---|---|---|
| Hydrodynamic Parameters | $K_s$ | 8.3E-06 | [m/s] | Saturated hydraulic conductivity |
|  | $\theta_r$ | 0.06 | [-/-] | Residual moisture content |
|  | n | 0.38 | [-/-] | Porosity |
|  | alpha | 0.93 | [1/m] | van Genuchten alpha parameter |
|  | N | 1.58 | [-] | van Genuchten N parameter |
| Vrugt et al., Root-water uptake model parameters | $z_m$ | 1 | [m] | Maximum rooting depth in soil profile |
|  | $r_m$ | 5 | [m] | Maximum rooting radius in soil profile |
|  | $z^*$ | 0 | [-] | Empirical parameter |
|  | $r^*$ | 0 | [-] | Empirical parameter |
|  | $p_z$ | 0 | [-] | Empirical parameter |
|  | $p_r$ | 0 | [-] | Empirical parameter |
|  | $h_{50}$ | -204 | [m] | Soil water pressure at which root water uptake reduced by 50% |
| Electrical Parameters | $\sigma_f$ | 0.0022 | [S/m] | Fluid electrical conductivity |
|  | $\sigma_s$ | 0.0006 | [S/m] | Surface conductivity |
|  | $\hat{Q}_v$ | See Figure 2 | [C/m³] | Effective excess charge, after Jougnot et al. (2012) |

## 5. Field Results

### 5.1. Precipitation

During calendar year 2016, HJA received at total of 2091 mm of precipitation, of which 474 mm fell during the period analyzed in this work, June 28 to November 3 (Figure 4a). A total of 88 mm fell in the early part of our data collection, between June 28 and September 30. During this dry period, most precipitation occurred during discrete events greater than 10 mm/day on July 8, July 10 and September 21. Smaller precipitation events (2–10 mm/day) occurred on August 17, September 1-2, 6-7, 17-18 and 22-23. In contrast to the relatively dry conditions during the summer months, 373 mm of precipitation occurred between October 1 and November 3 (the end of data collection). A total of 41 mm of precipitation fell on October 4 and 5 marking the transition to the wet period.

### 5.2. Tree Transpiration

Due to technical malfunctions, sap flow measurements were not obtained July 9 - 14, 2016 (Figure 4a). The data gap was filled using sap flow rates from sensors in eight nearby (<100m) Douglas-fir trees. The average sap flow values of the surrounding trees were normalized to the sap flow measurements recorded on the selected Douglas-fir tree during the ten days before and ten days after the data gap. The following observations are based on transpiration rate calculations (detailed in section 3.3) from the corrected sap flow data (Figures 4a, 5a and 5f).

During the period of measurement, daily total tree transpiration rates were greatest in June and July and decreased throughout the summer (Figure 4a). Negligible transpiration occurred on July $8^{th}$ and July $10^{th}$ due to low levels of solar radiation and low vapor pressure deficit during precipitation events. Thirty-minute data from July 5-8 and September 8-12 are shown in Figures 5a and 5f. Within a day, the highest rates of transpiration occurred during midday hours, and returned to near-zero overnight.

*5.3. Soil Moisture*

As with precipitation, the soil moisture data can also be divided into two periods: dry and wet (Figure 4c), with the separation between the two also occurring in early October. During the dry period from July through September, precipitation was low and the soil moisture measured at 0.3 m and 0.8 m depths steadily decreased (Figure 4c). Soil moisture at 0.8 m dropped from 0.13 $m^3/m^3$ at the beginning of July to 0.09 $m^3/m^3$ at the end of September. During the dry period, the soil moisture at the 0.3-m sensor was consistently 0.03 $m^3/m^3$ lower than the soil moisture at 0.8 m.

The soil moisture at 0.3 m abruptly increased from 0.06 $m^3/m^3$ to 0.17 $m^3/m^3$ on October 5 after the onset of rainfall described above (Figure 4c). Comparable increases in soil moisture at 0.8 m were observed nine days later on October 14. From October 14 until November 3, soil moisture at both depths remained above 0.17 $m^3/m^3$. A maximum soil moisture content of 0.22 $m^3/m^3$ occurs at 0.3 m on October 20. During the wet period, soil moisture at both depths increased in response to precipitation events and then gradually decreased. The magnitude of these short-term soil-moisture increases was greatest at 0.3 m, with more muted responses at 0.8 m. No detectable diel variations in soil moisture occur in the soil moisture records at either depth.

*5.4. Matric Potential*

The transition from the dry to wet period in early October was present in the matric potential data as a rapid increase in pressure (Figure 4d). During the dry period, the matric potentials measured at 0.3 m and 0.8 m depth gradually became more negative. During early

July, August and September, the matric pressure at 0.3 m depth was more negative than the matric pressure at 0.8 m. Diel fluctuations were not detectable in matric potential at 0.3 m for the period of July 5-9 (Figure 5d and 5i), but were weakly present during the period of September 8-11 when matric potentials are more negative (Figure 5i). Within the diel fluctuations, minimum matric potential occurred coincident with the maximum of transpiration (Figure 5f). No diel fluctuations in matric potential at 0.8 m depth were observed (Figures 5d and 5i).

### 5.5. *Field SP Data*

Temperature-corrected SP data (Equation 7), in mV, for all four electrode pairs through the data collection period are shown in Figure 4e. As explained in the methods section, the pairs are referred to as the inner (0.1 m from the tree), outer (0.9 m from the tree), upper (0.3 m below the ground surface) and lower (0.8 m below the ground surface) pair corresponding to their positions relative to the tree and ground surface (Figure 3a). Given the complexity of SP signal generation, and the resulting interpretation, we discuss only changes in mV which occur in the signals in response to specific events, and reserve interpretation of underlying direction of soil-water movement for the Discussion section.

#### 5.5.1. Discrete Events

Voltages from the four electrode pairs reveal two distinct periods, distinguished primarily by a change in signal magnitude, defined as the absolute value of the measured voltage differences (Figure 4e). Signal directions, or polarity, remain constant during the measurement period, except where noted below. The signal magnitude of all electrode pairs increased in early October, coinciding with the onset of the wet period as also recorded by precipitation, saturation and matric potential. During the dry period, voltages for all electrode pairs were smaller ($< \pm 40$ mV) than during the wet period ($> \pm 40$ mV), and the absolute value of the voltage differences of the inner and lower pairs remained near zero, except for minor deviations (<10 mV) on July 8-10 and August 17, which were smaller in magnitude than the upper and outer pairs. During the dry period, the polarity of the inner pair changed from positive to negative.

During the dry period, the measured voltages changed in response to precipitation events, but the timing and magnitude of the responses was not linearly related to the precipitation rate or total. Signal magnitudes of the outer and upper pairs increased in response to all three dry-season precipitation events greater than 10 mm (Figure 4e, discussion of precipitation in section 5.1). However, immediately after the onset of the July 8 precipitation event, the voltage difference in the upper pair decreased by 1.5 mV, prior to the net increase of 10 mV. While each of the precipitation events greater than 10 mm resulted in a net increase in signal magnitudes in the outer and upper pair, the timing varied. Whereas the signal magnitude increases of July 8 and September 21 occur within 24 hours after the onset of the precipitation events (Figures 4e and 5e), it appears that the signal increases in response to the July 10 precipitation do not occur until July 15. The response of the signal magnitudes of inner and lower pairs were not equal.

The signal response to smaller precipitation events is inconsistent. A small precipitation event on August 17 (2 mm) resulted in sharp decreases in signal magnitude in both the outer and upper pairs, whereas precipitation from September 6-7 (7 mm) and September 17-18 (10 mm) resulted in an increased magnitude of the outer pair (more negative), while the magnitudes of the upper pair decreased. Changes in the inner and lower voltage differences in response to the August 17 event do not appear until five days later (Figure 4e).

Finally, a jump of 53 mV occurs on October 29 in only the inner and lower electrode pairs, with no observable change in the outer and upper pairs. This is thought to have resulted from technical, rather than environmental reasons (see Discussion).

5.5.2. Diel Fluctuations

Diel fluctuations in voltage differences were present in some portions of the SP data. In July (Figure 5e), diel fluctuations were most prominent in the vertical (outer and inner) electrode pairs, with changes up to 1 mV occurring over the course of a day. The voltage differences were highest during the day, and decreased overnight. On July 8, when precipitation began and there was no measurable transpiration (Figure 5a), the diurnal increase in voltage differences did not occur (Figure 5e).

Unfortunately, the diel signals from mid-July onward were obscured in part by unexpected electrical noise created by a solar panel added to the data logger during data collection. Despite this setback, indications of diel variations were still present in all electrode pairs (Figure 5j), but are most pronounced, like before, in the vertical inner and outer pairs as compared to the horizontal electrode pairs (upper and lower). In the outer pair, the amplitude of diel fluctuations increases through the summer; fluctuations of almost 1.5 mV occur in late August.

*5.6. Modeling Results*

Before presenting the modeled SP signals, it is important to note that we do not use the model to quantify the flow rates that produced the observed signal using this model, nor do we match the hydrologic data exactly. The differences between our simple model and the field data stem from the numerous, poorly constrained parameters required in modeling unsaturated flow processes. Additionally, there is the possibility that our model of porous media flow does not include processes present at the site, such as macropore flow. Despite these differences, we feel that the general flow patterns produced by the hydrologic model are suitable for evaluating whether root-water uptake could produce the SP signals we observe. The shape and timing of the SP curves, rather than the absolute values of the signals of interest. In this manner, we demonstrate that SP is a tool that can be used to analyze root-water uptake dynamics at this spatial scale.

Figures 6e, 7e and 7j show the simulated voltage differences from the coupled model over the same data-collection time periods as Figures 4e, 5e and 5j. The modeled voltage differences increased in magnitude between the wet and dry period as seen in the field data. However, the modeled inner and lower pairs deviated in the opposite direction from their measured counterparts during the wet season. Instead of becoming more negative during the wet period, the modeled SP signal of the lower electrode pair became slightly more positive.

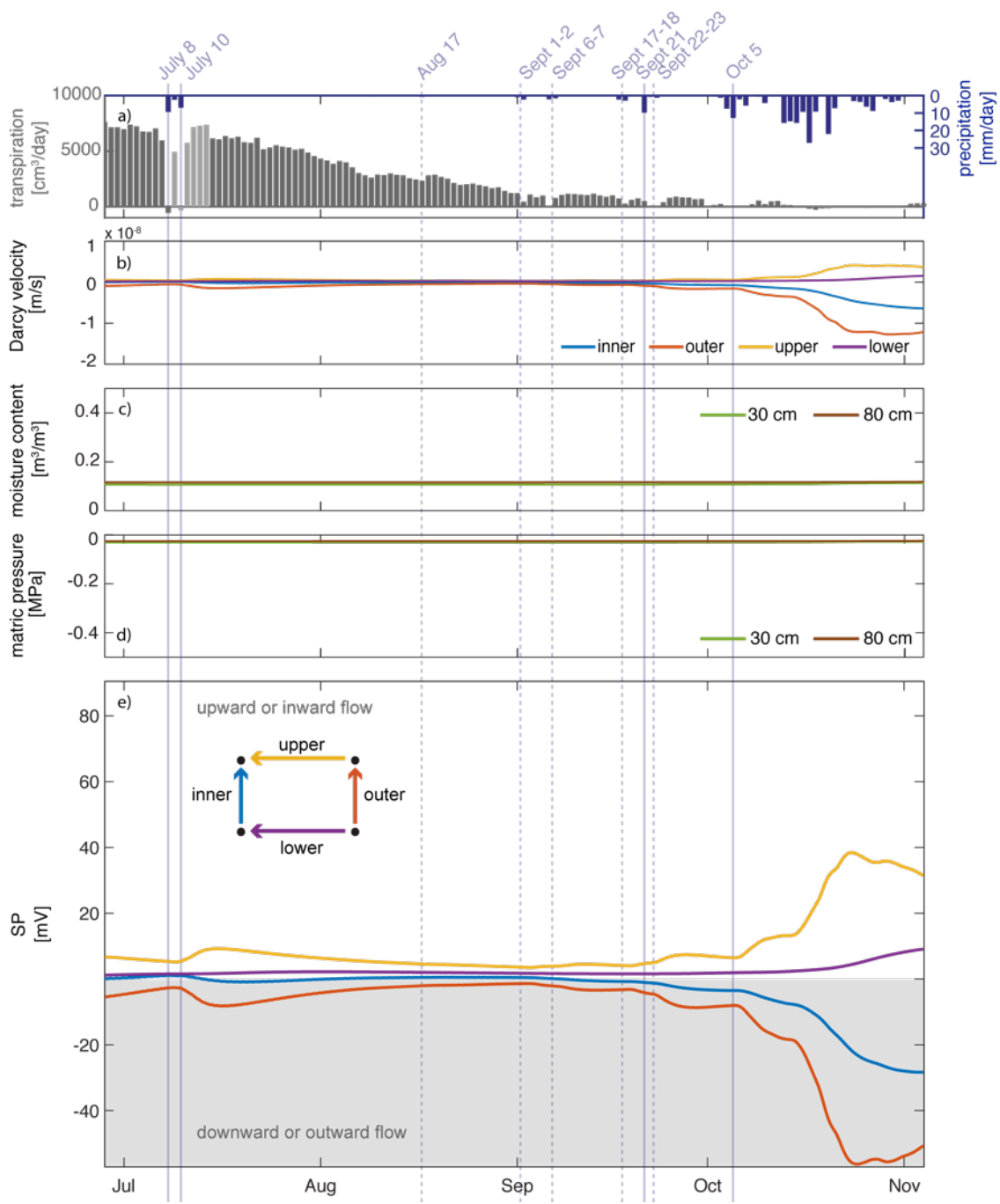

*Figure 6. a) Tree transpiration calculated from measured sap flow and precipitation used as boundary conditions in the model, b) simulated Darcy velocity, c) simulated soil moisture, d) simulated matric head, and e) simulated SP voltages between subsurface electrodes.*

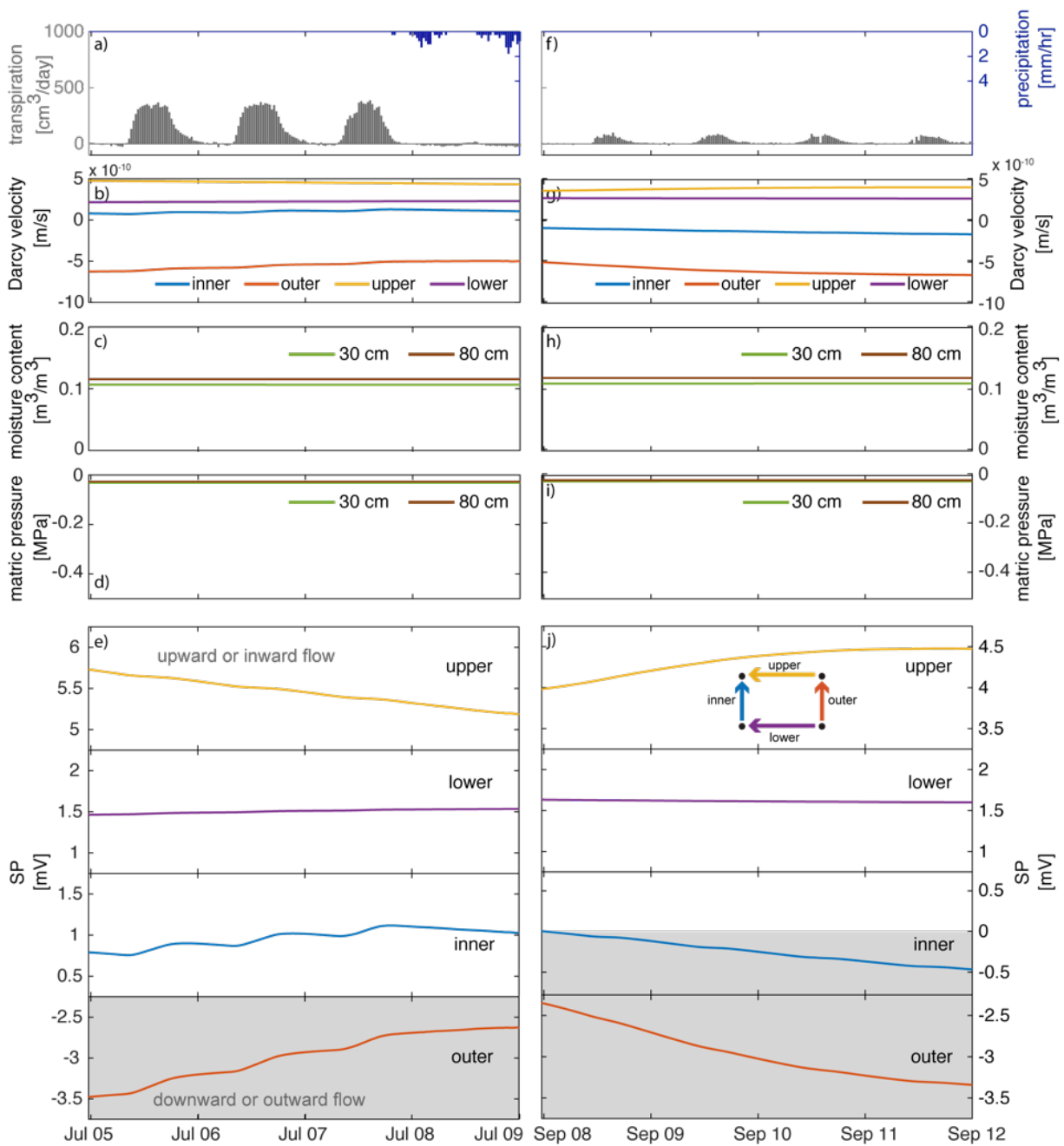

*Figure 7. Simulated data over four days: a-e) July 5-8 and g-j) September 8-11. a) and f) Tree transpiration calculated from measured sap flow and precipitation recorded at the HJA PriMet station, b) and g) simulated Darcy velocity, c) and h) simulated soil moisture, d) and i) simulated matric potential, and e) and j) simulated measured SP voltage differences between each pair of subsurface electrodes.*

The simulated voltages in all electrode pairs increased in magnitude in response to the >10 mm precipitation events of July 8, July 10 and September 21, similar to the field data. After the precipitation, the voltages returned to pre-event levels, although more slowly than in the measured data. Not surprisingly, the modeled data were smoother than the measured data, and did not contain some of the jumps observed in the measured data. On a daily timescale, the modeled SP signals contained diel fluctuations in the July excerpt, especially in the inner and outer pair (Figure 7e), but at a much smaller amplitude than those observed in the measured data (Figure 5e).

6. **Discussion**

   *6.1. Measured SP data*

Assuming uniform flow conditions and the given sign convention of the SP data (Section 3.2), the consistently positive voltage difference of the upper electrode pair suggests that flow moved towards the tree around 0.3 m soil depth (Figure 5e). In contrast, the voltage difference between the lower pair was initially slightly positive, but then became increasingly negative throughout the measurement period, suggesting that movement at 0.8 m depth was away from the tree (downslope) during the wet period. The signal recorded in the vertical pairs also suggests opposing movement directions, particularly during the wet period. The voltage of the inner pair, closest to the axis of the tree (Figure 3c), was predominantly positive during the wet period, suggesting upward flow occurred below the axis of the tree (Figure 5e). In contrast, the voltage of the outer pair was negative during the wet period, suggesting that downward flow occurred farther away from the tree axis.

One explanation for opposing flow directions of parallel electrode pairs, as interpreted from the SP data, is the position of the electrode pairs relative to the root zone of the tree (Figure 3c). Water removal by tree uptake results in more negative matric pressures, potentially resulting in water movement towards the roots. The root density of Douglas-fir trees is inversely related to the depth and distance from the stem (Curt et al., 2001; Mauer and Palátová, 2012). Therefore, the inner and upper pairs, which both used the shallow electrode nearest the tree trunk, would be most sensitive to water movement within the root zone. In contrast, the outer and lower pairs, which are farther from the tree trunk, could be sensitive to flow occurring downward (infiltration) and away from the tree stem (downslope) at these locations. A lateral component of unsaturated hillslope flow after rainfall has been previous observed in this watershed (Harr, 1977; van Verseveld et al., 2017). These interpretations will be addressed more in the discussion of the measured and modeled data (Section 6.3).

*6.2. Direction of water movement*

Rather than supporting a common interpretation of flow direction, the SP and tensiometer data suggest contrasting directions of flow. The SP data from the outer pair suggest downward soil movement throughout the entire measurement period (Figure 4c). In contrast, the potential gradient (not shown, individual matric potential values are in Figure 4d), indicates conditions for upward water movement during the dry season, which shifts to downward movement only after the start of the wet period. Soil moisture sensors indicate an increase in soil moisture at the 0.3-m and 0.8-m sensors following discrete precipitation events during the dry period (Figure 5c, Jul 8), which implies that downward movement of water during this time did occur. One possible explanation for the difference in interpretation is that uncertainty in the matric potential measurements could make them unsuitable for differential measurements at this scale (0.5-m separation). The manufacturer-reported accuracy of these sensors scales as 5% of the measured value (i.e., logarithm of pressure), with a minimum value of ±0.003 MPa or 0.3 m of water pressure head. Consequently, as matric potential increases, there is increasing error as the soil dries out. At the maximum pressure recorded in our study, 1.3 MPa, this error equivalent to ±0.5 MPa.. This error is typical for this type ceramic tip heat dissipation matric potential sensor (Phene et al., 1971), and could be a reason that hydraulic gradients calculated from tensiometers at this spacing are rarely reported in the literature.

Another possibility for the difference in interpretation is the uncertainty of the SP measurements. We have assumed that the polarity of the SP signals is indicative of the direction of water movement; however, while SP is sensitive to the movement of water, it is sensitive to other processes (e.g., Revil and Jardani, 2013) and influenced by three-dimensional flow (Voytek et al., 2016). Heterogeneity of the soils on site likely invalidate the assumption of uniform 1-D or 2-D flow. In addition, SP measurements are dependent on the electrical conductivity structure of the subsurface (Equations 14 and 15) and the saturation-dependent effective excess charge (Equations 11 and 15). Therefore, signal measured by SP may not always be coincident with the direction of fluid flow. However, the model results, discussed below, support the directional interpretations of the SP data from the outer electrode pair, in which the negative voltage differences measured consistently in the outer pair of electrodes are produced by constant downward movement of fluid, rather than upward movement as indicated by the tensiometers.

### 6.3. Measured vs. modeled SP data

The goal of the numerical modeling exercise was to explore the mechanisms controlling the observed signals rather than to duplicate signals exactly; therefore, the modeled values were not identical to the field data, and in some instances were substantially different than the field data. The complex origins of the SP signals, resulting from interaction of multiple hydraulic and electrical parameters, precludes developing a model with complete internal consistency to the field data. The primary differences between the measured data and the model were: 1) polarity of the outer and lower pairs' voltages, 2) variations in exact values in those voltage differences, 3) fit of the hydrologic model and 4) the overall smoothness of the modeled SP data.

While the modeled inner and upper pairs had the same general patterns as the measured data—low values increasing with the onset of wet season—the values of the lower and outer pairs in the modeled data (Figure 4e) deviated in opposite directions from their measured counterparts (Figure 6e) during the wet season. Instead of becoming more negative during the wet period as in the measured data, the modeled SP signal of the lower electrode pair becomes more positive. The lack of the 37° hillslope present at the field site in our modeled system may explain these differences. Because we primarily used the model to test the origin of signals, rather than match the signals exactly, the hillslope was not included, as it substantially increased the complexity of an already uncertain model. As a result, in the model, precipitation results in water flow toward the low soil moisture area created by root-water uptake at both 0.3 m and 0.8 m depths. There is no component of downslope flow. This results in positive voltages in both the upper and lower pairs, and negative in the inner and outer pairs (Figure 6e). However, in the field data, the lower electrode pair could be influenced by topographically driven downslope flow, and result in the negative voltages measured in the lower electrode pair.

Another notable difference between the measured and modeled data are the hydrologic model results (e.g. moisture content and matric potential). We estimated the hydrologic model parameter values from available field data and previously published values (Table 1). The model did not produce the large negative matric potentials that were measured in the field. This difference may be due to uncertainty of the soil-moisture release curve, or could also derive from the uncertainty of the transpiration calculations, which determined the quantity of water extracted in the model. The differences in the measured and modeled hydrologic values may also contribute to the differences in the modeled SP values.

The simulated SP data differ in magnitude, and at times, in the relative direction of signal change. For example, in the July excerpt the measured upper and outer voltage differences maintained a steady background value despite diel fluctuations (Figure 5e), while in the numerical model, the background values gradually decreased and increased respectively (Figure 7e). As with the hydrologic simulations, differences in the shape and slope of the measured and modeled SP values derive from the uncertainty in a number of parameters used in the model. The interrelated values of the parameters (e.g., van Genuchten parameters with $\hat{Q}_v$ (Equation 11), and soil moisture with resistivity (Equation 14)), make it nearly impossible to match every part of the dataset. Given the uncertainty associated with many parameters used in the model, it promising that the modeled values are within an order of magnitude to those observed in the field (Figure 5e vs. Figure 7e).

Finally, as expected for a simplified model, the modeled data were smoother than the measured data. The modeled data do not contain all of the details observed in the measured data. While our model produced gradual increases in voltage differences in response to precipitation events in July and September similar to those observed the measured data, it did not reproduce some of the more abrupt jumps. For example, the jumps that occur on July 8, August 17, August 21 and September 22 are not present in the modeled data. This suggests our model does not include some processes contributing to the SP signals. In additional to water flow, as noted earlier, SP measurements are also sensitive to thermal, chemical and redox gradients (Revil and Jardani, 2013). These processes are assumed to have generally have a minor contribution in this setting and thus were not included in our model.

### *6.4. SP response to precipitation*

Voltage differences between all electrode pairs increase in amplitude from the dry to wet period. Increases in voltage difference magnitude can be caused by increasing the quantity of soil water (i.e. saturation) or increases in the velocity of soil-water flow. The increases observed in our data likely result from both mechanisms. Increases in moisture content are recorded at both 0.3 m and 0.8 m after the start of the wet period (Figure 4c), and in the model increased precipitation results in increased rate of soil moisture movement (Figure 6b). Increases in SP voltages, followed by a gradual decline to baseline values, have been observed in association with precipitation events by Doussan et al. (2002) and modeled by Darnet and Marquis (2004). However, not all precipitation events during our measurement period result in increased voltages followed by a steady decline. The precipitation on September 21 results in immediate increases in the upper and outer voltage differences, but eventual increases in the inner and lower voltage differences are proceeded by abrupt decreases. A similar abrupt decrease, proceeding an eventual increase, occurs in the upper pair following the July 8 precipitation. These are followed by more gradual increases of the upper- and outer-pair magnitudes, but not until July 15, after additional precipitation occurs on July 10. Polarity changes in SP data have been observed in association with the onset of head changes from pumping tests (Malama, 2014). Precipitation events, which initiate quick changes in water content and matric potentials, may have similar effect, although it is clear that this effect is not linear to the expected changes in those properties: the greatest drop in voltage magnitude occurs on August 17 following only 2 mm of precipitation. Changes in the inner and lower voltages, presumably in response to the August 17 event, do not appear until three days later (Figure 4e).

The differing response of the SP signals to precipitation events could be related to the state of the system (e.g. soil moisture level) prior to precipitation occurring. For example, on July 8, when soil moisture is slightly higher, the voltage differences between each electrode pair responds within a few hours (end of time series in Figure 5e), whereas in mid-August when conditions are only slightly drier, the response of the inner and lower pairs is delayed by three days. Possible time delays of responses, combined with multiple precipitation event in August complicate the analysis of the individual precipitation events, as the responses may be overlapping.

Lastly, we explore a discrete jump of 53 mV on October 29 in only the inner and lower electrode pairs. This occurs after the onset of the wet period, and there is no corresponding change in the outer and upper pairs. The features of this jump, including a change of exactly the same voltage in two pairs sharing a common electrode (Figure 3c) and no changes in the other pairs, suggest that it is caused due to an electrode malfunction rather than environmental change. The differing polarity of the jump—one positive and one negative—is due to the relative orientation of the shared electrode in the pair (Figure 3c). The exact cause of this jump is unknown, but could be due to oxidation at the wire tip inserted in the data logger, or deterioration in the electrode-soil contact. Poor electrode-soil contact typically occurs under dry conditions (e.g., Doussan et al., 2002), rather than during wet conditions as observed here, so this hypothesis seems unlikely. No sudden voltage jump occurs in the modeled data on October 29, further supporting the interpretation of a technical malfunction, rather than environmental origin.

### *6.5. Daily variations*

As stated in Section 5.3, no diel signals were present in the soil moisture data collected during the study period. In contrast, Barnard et al. (2010) observed diel fluctuations of soil moisture of up to 10% within the same watershed at HJA during a 4-day period in July 2005. This discrepancy could be due to the use of different sensors (Environmental Sensors, Inc., model PRB-A vs Decagon, model EC-5), different hydrologic conditions (possible differences in soil moisture) or even the difference in placement of the sensors within the watershed. The exact values of soil moisture values measured in these two studies cannot be used to compare the relative effects of soil moisture differences, because neither the sensors used in Barnard et al. (2010), nor those used in this study, were calibrated for the site-specific conditions.

Despite a lack of diel signals in the soil moisture data, diel variations are present in some portions of the SP data. Two excerpts from July 5-9 (Figure 5e) and September 8-12 (Figure 5j) have been included as examples. During both periods, the outer pair voltages, which were consistently negative (indicative of downward flow), became less negative during the day (indicative of decreased downward flow), and more negative (indicative of increased downward flow) overnight. The decrease in amplitudes of measured voltages coincides with the peak in transpiration. Similar behavior occurs in the inner pair, despite possible uncertainties of absolute direction of flow (Section 6.2). The positive values become more positive during the day, suggesting movement is towards the ground surface (tree axis).

Diel variations in SP associated with tree transpiration have been recorded before. Gibert et al. (2006) continuously measured SP in electrodes installed into a tree trunk and the surrounding roots to quantify variability of sap flow within a tree. In addition to the multiple trunk and root electrodes, Gibert et al. (2006) measured the potential between the single reference electrode and a second ground electrode (called a "soil chemical electrode" in the paper) closer to the tree. The diel potential differences measured between the two ground electrodes (comparable to our upper horizonal pair of our data) were hypothesized to be related to changes in soil chemistry, but no corroboratory data were presented. It is possible that the diel changes are signals are due to transpiration, as considered here. The potential differences are positive (suggesting flow towards the tree) and increase during transpiration, suggesting movement towards the tree controlled by transpiration. However, the diel fluctuations are only detectable during precipitation-free periods, because precipitation events produce larger magnitude signals (Section 6.4).

## 7. Conclusions

Here, we collected SP data to analyze two-dimensional water movement in association with root-water uptake of a Douglas-fir tree in the H.J. Andrews Experimental Forest in Oregon over four months. In particular, we investigated sensitivity of SP measurements to evaluate changes in subsurface water flow induced by precipitation and transpiration. Continuous soil moisture, matric potential, temperature and SP measurements, in combination with a coupled fluid flow and electrical modeling, were used to evaluate the measured SP signals from the field, and investigate the propagation of transpiration and precipitation signals in the subsurface.

Daytime increases in sapflow correspond with reduced rates of downward flow in SP measurements that are not detectable as changes in water content. SP and tensiometer data were interpreted to show contrasting directions of flow during the dry season, due to uncertainty in the tensiometer or SP data. Given the directional discrepancy between the SP data and the tensiometer data, we do not definitively state the direction of water movement in this system. Instead, we observe that the SP measurements respond to precipitation events and diel changes in transpiration. The diel signals disappear when transpiration was absent, or when precipitation-induced signals are present.

The goal of the modeling exercise was to explore the mechanisms controlling the observed signals rather than to duplicate signals exactly given the uncertainties of the interrelated electrical and hydrological parameters; therefore, the magnitudes of the modeled SP signals are not identical to the field data. Even though the modeled matric potentials vary from the field observations due to uncertainty of soil moisture and porosity parameters, the model does confirm the transpiration origin of the SP signals. Additionally, the model supports the ability to this method to detect subsurface processes of this scale. With additional exploration of SP signals, this method could be useful for evaluating patterns of flow at this scale. However, the lingering sources of uncertainty in the methods outlined here should be addressed prior to data collection in other systems.

## 8. Acknowledgements

Data and facilities were provided by the HJ Andrews Experimental Forest and Long Term Ecological Research program, administered cooperatively by the USDA Forest Service Pacific Northwest Research Station, Oregon State University, and the Willamette National Forest. This material is based upon work supported by the National Science Foundation under Grant No. DEB-1440409 and EAR-1446161, EAR-1446231. This work was conducted with U.S. Government support awarded to EBV by the DoD Air Force Office of Scientific Research National Defense Science and Engineering Graduate (NDSEG) Fellowship, 32 CFR 168a. Additional support to EBV was provided by the American Geophysical Union Horton Research Grant. We thank Jackie Randell, Ryan Harmon and Daphne Szutu for field assistance, Stephanie Jarvis for transpiration calculations, Andre Revil, Niklas Linde and Michelle Walvoord for extended discussions, and Rania Eldam Pommer for tree illustrations. We thank the editor and three anonymous reviewers whose detailed comments greatly improved the manuscript.